\documentclass{mn2e}

\def\spose#1{\hbox to 0pt{#1\hss}}
\def\lta{\mathrel{\spose{\lower 3pt\hbox{$\mathchar"218$}}
     \raise 2.0pt\hbox{$\mathchar"13C$}}}
\def\gta{\mathrel{\spose{\lower 3pt\hbox{$\mathchar"218$}}
     \raise 2.0pt\hbox{$\mathchar"13E$}}}

\def\FMG{{F_{\rm MgII{\it hk}}}}
\def\FMGL{\log{F_{\rm MgII{\it hk}}}}

\def\1p{\phantom{0}}
\def\2p{\phantom{00}}
\def\3p{\phantom{000}}
\def\4p{\phantom{0000}}
\def\ithk{{\it h}+{\it k}}

\usepackage{graphics,epsfig}

\title[Basal chromospheric flux of cool giant stars]
{The basal chromospheric Mg~II {\ithk} flux of evolved stars:
Probing the energy dissipation of giant chromospheres}
\author[M. Isabel P{\'e}rez Mart{\'{\i}}nez, K.-P. Schr\"oder, M. Cuntz]
{M. Isabel P{\'e}rez Mart{\'{\i}}nez$^{1}$,  
K.-P. Schr\"oder$^{1}$\thanks{email: kps@astro.ugto.mx},
and M. Cuntz$^{2}$ \\
$^{1}$Departamento de Astronomia, Universidad de Guanajuato, 
A.P. 144, Guanajuato, GTO, C.P. 36000, Mexico \\
$^{2}$Department of Physics, University of Texas at Arlington, Box 19059,
Arlington, TX 76019, USA
}

\begin{document}

\date{Accepted 2011 .....; Received 18.07. 2010}


\maketitle

\begin{abstract}
Of a total of 177 cool G, K, and M giants and supergiants, we measured 
the Mg~II {\ithk} line emission of extended chromospheres in 
high-resolution (LWR) IUE spectra by using the IUE final data archive 
at STScI, and derived the respective stellar surface fluxes. 
They represent the chromospheric radiative energy losses presumably
related to basal heating by the dissipation of acoustic waves, 
plus a highly variable contribution due to magnetic activity.

Thanks to the large sample size, we find a very well-defined 
lower limit, the basal chromospheric Mg~II {\ithk} line flux of cool giant 
chromospheres, as a function of $T_{\rm eff}$.
A total of 16 giants were observed several times, over a period of up
to 20 years.  Their respective minimal Mg II {\ithk} line fluxes confirm
the basal flux limit very well because none of their emissions dip
beneath the empirically deduced basal flux line representative for the
overall sample.  Based on a total of 15 to 22 objects with very low
Mg~II {\ithk} emission, we find as limit:
$\FMGL = 7.33 \log{T_{\rm eff}} - 21.75$ (cgs units; based on the $B-V$ relation).
Within its uncertainties, this is almost the same relation as has been found 
in the past for the geometrically much thinner chromospheres of main sequence 
stars. But any residual dependence of the basal flux on the surface 
gravity is difficult to determine, since especially among the G-type giants 
there is a large spread of the individual chromospheric Mg~II fluxes, 
apparently due to revived magnetic activity. However, it can be stated 
that over a gravity range of more than four orders of magnitude 
(main-sequence stars to supergiants), the basal flux does not appear to vary by 
more than a factor of 2. 

These findings are in good agreement with the predictions by previous 
hydrodynamic models of acoustic wave propagation and energy dissipation, 
as well as with earlier empirical determinations. Finally, we also discuss 
the idea that the ample energy flux of the chromospheric acoustic waves 
in a cool giant may yield, as a by-product, the energy flux required by 
its cool wind (i.e., non-dust-driven, ``Reimers-type'' mass-loss), provided a
dissipation mechanism of a sufficiently long range is operating.
\end{abstract}



\section{Introduction}

The physical processes operating within the chromospheres of cool stars
are very complex and are still far from being fully understood. 
The dissipation of chromospheric energy can be probed by the related radiative 
cooling via chromospheric line emission, most importantly the Mg~II and 
Ca~II lines, see Linsky \& Ayres (1978).
As discussed in more detail by Judge (1990), radiative cooling is
not much affected by the optical thickness of the lines. The photons 
leave from a low-density giant chromosphere after multiple 
line-scattering (``effectively thin''). Empirically, the chromospheric 
line emission shows a minimal, i.e., ``basal'' flux level, that depends on 
the effective temperature (Strassmeier et al. 1994; Schrijver 1987; 
Rutten et al. 1991), above which there is a wide distribution of 
line emissions for the different stars (Vilhu \& Walter 1987),
which also may vary with time. 

It has been suggested to associate the basal flux with the dissipation 
of purely mechanical (i.e., acoustic wave) energy in the turbulent 
chromosphere (e.g., Buchholz, Ulmschneider \& Cuntz 1998), while energy released by 
magnetic activity phenomena is expected to produce the wide variation of emission 
beyond the basal flux. 
Albeit this is a simple picture, there is still a number of open questions
remaining.  Nevertheless, this view appears to be 
at least consistent with existing empirical evidence for
chromospheric energy dissipation (e.g., Ulmschneider 1991;
Cuntz, Ulmschneider \& Musielak 1998; Cuntz et al. 1999;
Theurer, Ulmschneider \& Kalkofen 1997; 
Fawzy et al. 2002; see also review by Musielak 2004). 
Specifically, a similar dependence on the effective temperature is obtained 
($\propto T_{\rm eff}^{7...8}$; see Musielak \& Rosner 1988 and Buchholz 
et al. 1998, particularly their Fig.~15) for both the initial acoustic
wave energy flux and the radiative emission flux observed as basal emission.
Furthermore, acoustic wave dissipation is mostly independent of gravity 
(see Ulmschneider 1988, 1989 and Cuntz, Rammacher \& Ulmschneider 1994), providing 
a further empirical prediction that can be tested by observation.

While earlier studies (e.g., Strassmeier et al. 1994) have already established 
that the basal flux emission declines towards cooler stars with a large power of 
$T_{\rm eff}$, this type of work was mostly based on main-sequence (MS) stars.
In contrast, our focus is on the basal flux emission of low-gravity stars;
i.e., stars with about two (giants LC~III) to over four orders of magnitude
(supergiants LC~I) lower surface gravity than the Sun.  Hence, we
not only aim at refining the dependence of the basal flux on $T_{\rm eff}$,
but also at contesting any possible dependence of the basal flux on gravity.

Giant stars also present an interesting case, noting that only a small fraction
of their chromospheric mechanical flux would be needed to satisfy the input requirements 
of their non-dust-driven (``Reimers-type'') cool winds, assuming that a mechanism with 
a sufficiently large dissipation length operates (e.g., 
Holzer \& MacGregor 1985; Cuntz 1990; Sutmann \& Cuntz 1995).
Schr\"oder \& Cuntz (2005) have shown, based on a simple analysis of the 
energy requirements of the winds, that such a mechanism would be well 
consistent with the Reimers formula (Reimers 1977). The resulting,
improved mass loss relation was then successfully tested by Schr\"oder \& 
Cuntz (2007). An updated concept of structured red giant chromospheres and
winds initiated by magnetized hot bubbles and the action of Alfv\'en waves, 
including long-range action by mode-coupling, has been given by Suzuki (2007).
His models are also consistent with the existence of a small hot gas component 
apparently buried under a large column of chromospheric material
(Ayres, Brown \& Harper 2003).

To revisit the empirical evidence, we measure the chromospheric emission 
line fluxes of 177 giant and supergiant stars (Sect. 2). Thereafter, 
we determine the basal emission flux for this sample, including
its dependence on $T_{\rm eff}$ (Sect. 3), by a detailed statistical
analysis. We also take a look at how well the observed variability of 
19 stars over periods of up to two decades is consistent with the basal 
flux concept. As we find no evidence for a significant dependency on gravity, we 
derive an upper limit to any residual change of basal flux over a gravity
range of 4 orders of magnitude. Associated aspects of stellar evolution are 
elucidated in Sect. 4. Finally, we compare our results with previous empirical
and theoretical findings (Sect. 5) and present our discussion and 
conclusions (Sect. 6).

\begin{figure}
\centering
\begin{tabular}{c}
\epsfig{file=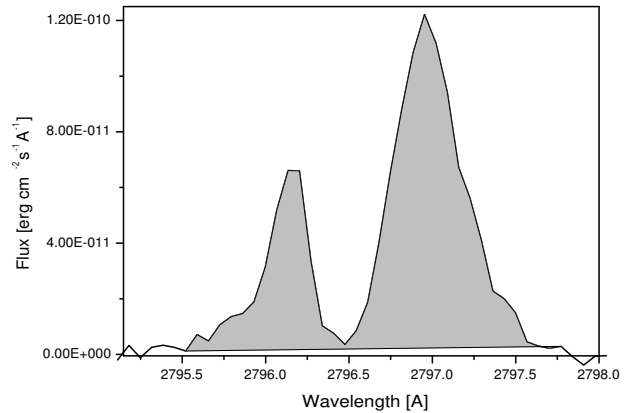,width=1.10\linewidth}
\end{tabular}
\caption{Mg~II {\it k} chromospheric emission line: the shaded area 
         shows the range considered for the flux integration.
}
\label{}
\end{figure}

\begin{table*}
\caption{Physical and empirical surface fluxes of the chromospheric 
Mg~II {\ithk} line emission of 190 cool giants and supergiants from the 
IUE final data archive, together with $B-V$, $V$, $BC$ and $\log{T_{\rm eff}}$ 
(see Sect. 2). Stars with a minimum flux obtained from several observations are 
marked by an asterisk ($\ast$). Stars used to determine the basal flux line using 
the $B-V$ color transformation are marked by a dagger ($\dagger$), whereas
stars associated with the $V-K$ color transformation are marked by a double dagger ($\ddag$).}
\begin{footnotesize}
\begin{tabular}{ l l l p{1.7cm} l l p{1.7cm} p{1.7cm} p{1.7cm} p{1.7cm}}\hline
HD	                       &	$B-V$	&	$V-K$	&	$\FMG$ 	&  $BC_{\rm V}$	&  $BC_{\rm K}$	& $\log{T_{\rm eff}}$ &	$\log{T_{\rm eff}}$ & $\FMGL$    & $\FMGL$	\\
...	                       &	...	&	...	&	...    	&  ...   	&  ...   	& from $B-V$          & from $V-K$          & from $B-V$ & from $V-K$	\\ \hline \hline
{\4p}28	                 &	1.02	&	2.59	&	4.06E-12	&	-0.40	&	2.17	&	3.682	&	3.657	&	5.37	&	5.25	\\
{\3p}352	                 &	1.27	&	3.02	&	1.18E-11	&	-0.71	&	2.36	&	3.640	&	3.627	&	6.17	&	6.03	\\
{\3p}496 $\dagger$	     &	1.00	&	2.22	&	6.43E-12	&	-0.38	&	1.94	&	3.685	&	3.690	&	5.30	&	5.34	\\
{\2p}1522	                 &	1.19	&	2.45	&	9.39E-12	&	-0.60	&	2.09	&	3.653	&	3.668	&	5.12	&	5.24	\\
{\2p}2261	                 &	1.08	&	2.54	&	3.80E-11	&	-0.47	&	2.14	&	3.672	&	3.660	&	5.39	&	5.36	\\
{\2p}3627	                 &	1.26	&	2.77	&	1.15E-11	&	-0.69	&	2.26	&	3.642	&	3.643	&	5.01	&	5.07	\\
{\2p}3712	                 &	1.15	&	2.48	&	3.03E-11	&	-0.55	&	2.11	&	3.660	&	3.665	&	5.15	&	5.22	\\
{\2p}4128	                 &	1.01	&	2.29	&	1.02E-10	&	-0.39	&	1.99	&	3.684	&	3.683	&	5.75	&	5.78	\\
{\2p}4174	                 &	1.40	&	3.99	&	3.69E-12	&	-0.92	&	2.65	&	3.618	&	3.586	&	5.89	&	5.35	\\
{\2p}4502	                 &	1.08	&	2.46	&	8.30E-11	&	-0.48	&	2.10	&	3.670	&	3.667	&	6.39	&	6.40	\\
{\2p}6805 $\dagger$	     &	1.15	&	2.51	&	8.70E-12	&	-0.55	&	2.12	&	3.660	&	3.663	&	5.09	&	5.16	\\
{\2p}6860	                 &	1.56	&	3.86	&	7.89E-11	&	-1.35	&	2.62	&	3.587	&	3.590	&	4.88	&	4.92	\\
{\2p}8512 $\dagger$	     &	1.05	&	2.28	&	7.59E-12	&	-0.45	&	1.98	&	3.675	&	3.684	&	5.19	&	5.27	\\
{\2p}9053	                 &	1.52	&	3.84	&	3.68E-11	&	-1.21	&	2.61	&	3.596	&	3.590	&	5.18	&	5.12	\\
{\2p}9746	                 &	1.20	&	3.17	&	3.75E-11	&	-0.61	&	2.42	&	3.652	&	3.619	&	6.76	&	6.53	\\
{\2p}9927	                 &	1.26	&	2.89	&	7.51E-12	&	-0.69	&	2.31	&	3.642	&	3.635	&	4.95	&	4.95	\\
{\1p}10380	                 &	1.31	&	2.99	&	5.98E-12	&	-0.78	&	2.35	&	3.632	&	3.629	&	5.12	&	5.13	\\
{\1p}12929	                 &	1.14	&	2.77	&	4.54E-11	&	-0.55	&	2.26	&	3.660	&	3.643	&	5.23	&	5.17	\\
{\1p}13480	                 &	0.74	&	2.19	&	2.37E-11	&	-0.15	&	1.92	&	3.736	&	3.693	&	6.58	&	6.33	\\
{\1p}17506	                 &	1.56	&	3.24	&	1.47E-11	&	-1.33	&	2.44	&	3.589	&	3.616	&	4.85	&	5.02	\\
{\1p}18322	                 &	1.07	&	2.38	&	8.02E-12	&	-0.47	&	2.05	&	3.672	&	3.674	&	5.31	&	5.36	\\
{\1p}18884	                 &	1.61	&	4.30	&	4.59E-11	&	-1.60	&	2.71	&	3.574	&	3.576	&	4.69	&	4.67	\\
{\1p}19476	                 &	0.97	&	2.52	&	8.08E-12	&	-0.35	&	2.13	&	3.690	&	3.662	&	5.39	&	5.25	\\
{\1p}20644 $\dagger$ $\ddag$ &	1.49	&	3.41	&	3.30E-12	&	-1.15	&	2.50	&	3.600	&	3.608	&	4.60	&	4.65	\\
{\1p}20720	                 &	1.59	&	4.75	&	2.68E-11	&	-1.49	&	2.79	&	3.579	&	3.565	&	4.98	&	4.71	\\
{\1p}23817 $\dagger$	     &	1.12	&	2.53	&	6.76E-12	&	-0.53	&	2.14	&	3.663	&	3.661	&	5.16	&	5.19	\\
{\1p}24512                   &	1.57	&	4.23	&	3.09E-11	&	-1.39	&	2.70	&	3.585	&	3.578	&	4.93	&	4.82	\\
{\1p}25025	                 &	1.57	&	3.86	&	2.72E-11	&	-1.39	&	2.62	&	3.585	&	3.590	&	4.75	&	4.81	\\
{\1p}26967	                 &	1.07	&	2.58	&	6.62E-12	&	-0.47	&	2.17	&	3.672	&	3.657	&	5.21	&	5.16	\\
{\1p}27371	                 &	0.97	&	2.09	&	1.03E-10	&	-0.35	&	1.84	&	3.690	&	3.705	&	6.44	&	6.52	\\
{\1p}27697	                 &	0.97	&	2.08	&	1.05E-11	&	-0.35	&	1.84	&	3.690	&	3.705	&	5.50	&	5.58	\\
{\1p}28305	                 &	1.00	&	2.06	&	9.69E-12	&	-0.38	&	1.82	&	3.685	&	3.708	&	5.34	&	5.47	\\
{\1p}28307	                 &	0.94	&	2.15	&	1.78E-11	&	-0.32	&	1.89	&	3.696	&	3.697	&	5.79	&	5.81	\\
{\1p}29139 *	           &	1.53	&	3.90	&	1.70E-10	&	-1.28	&	2.63	&	3.592	&	3.589	&	4.79	&	4.77	\\
{\1p}31398 *	           &	1.44	&	3.39	&	3.06E-11	&	-0.94	&	2.49	&	3.616	&	3.609	&	5.00	&	4.93	\\
{\1p}31767 $\dagger$ $\ddag$ &	1.27	&	2.86	&	3.42E-12	&	-0.66	&	2.30	&	3.646	&	3.637	&	4.99	&	4.89	\\
{\1p}32068	                 &	1.08	&	3.26	&	6.14E-11	&	-0.40	&	2.45	&	3.682	&	3.615	&	6.18	&	5.66	\\
{\1p}32887	                 &	1.44	&	3.33	&	1.70E-11	&	-1.00	&	2.47	&	3.611	&	3.611	&	4.90	&	4.93	\\
{\1p}32918	                 &	0.98	&	2.51	&	4.87E-12	&	-0.37	&	2.12	&	3.687	&	3.663	&	6.86	&	6.65	\\
{\1p}37160	                 &	0.94	&	2.25	&	9.57E-12	&	-0.32	&	1.96	&	3.696	&	3.687	&	5.62	&	5.59	\\
{\1p}39364	                 &	0.97	&	2.32	&	1.27E-11	&	-0.36	&	2.01	&	3.689	&	3.680	&	5.57	&	5.54	\\
{\1p}39425	                 &	1.14	&	2.41	&	1.50E-11	&	-0.54	&	2.06	&	3.662	&	3.672	&	5.20	&	5.32	\\
{\1p}39801 * $\ddag$	     &	1.46	&	4.71	&	3.02E-10	&	-0.99	&	2.78	&	3.612	&	3.565	&	5.06	&	4.45	\\
{\1p}40239	                 &	1.62	&	4.88	&	1.39E-11	&	-1.66	&	2.81	&	3.570	&	3.562	&	4.83	&	4.54	\\
{\1p}40409	                 &	1.01	&	2.32	&	3.11E-12	&	-0.40	&	2.01	&	3.682	&	3.680	&	5.27	&	5.29	\\
{\1p}42995	                 &	1.57	&	4.94	&	4.48E-11	&	-1.39	&	2.82	&	3.585	&	3.560	&	5.11	&	4.68	\\
{\1p}43039	                 &	1.00	&	2.56	&	5.89E-12	&	-0.39	&	2.15	&	3.684	&	3.659	&	5.43	&	5.30	\\
{\1p}44478 $\ddag$	     &	1.60	&	4.67	&	3.22E-11	&	-1.54	&	2.78	&	3.577	&	3.566	&	4.70	&	4.49	\\
{\1p}46697	                 &	1.09	&	2.54	&	2.41E-12	&	-0.48	&	2.14	&	3.670	&	3.660	&	6.28	&	6.21	\\
{\1p}47205	                 &	1.03	&	2.37	&	9.01E-12	&	-0.43	&	2.04	&	3.678	&	3.675	&	5.43	&	5.45	\\
{\1p}50310 $\dagger$         &	1.19	&	2.55	&	1.50E-11	&	-0.60	&	2.15	&	3.653	&	3.659	&	5.08	&	5.16	\\
{\1p}50877	                 &	1.54	&	3.03	&	9.78E-12	&	-1.28	&	2.37	&	3.591	&	3.627	&	4.75	&	4.92	\\
{\1p}52877	                 &	1.61	&	3.57	&	3.14E-11	&	-1.60	&	2.54	&	3.574	&	3.601	&	4.90	&	5.10	\\
{\1p}54810	                 &	1.00	&	2.41	&	3.77E-12	&	-0.38	&	2.06	&	3.685	&	3.672	&	5.48	&	5.42	\\
{\1p}56855	                 &	1.51	&	3.40	&	3.81E-11	&	-1.13	&	2.49	&	3.601	&	3.608	&	4.97	&	4.96	\\
{\1p}57669	                 &	1.18	&	2.60	&	1.71E-11	&	-0.59	&	2.18	&	3.655	&	3.655	&	6.06	&	6.04	\\
{\1p}59693 *	           &	0.87	&	2.14	&	5.56E-12	&	-0.21	&	1.88	&	3.719	&	3.699	&	6.61	&	6.26	\\
{\1p}59717	                 &	1.49	&	3.48	&	1.30E-11	&	-1.15	&	2.52	&	3.600	&	3.605	&	4.70	&	4.78	\\
{\1p}60414	                 &	1.03	&	3.82	&	3.93E-10	&	-0.36	&	2.61	&	3.689	&	3.591	&	7.48	&	6.38	\\
{\1p}61772 $\dagger$ $\ddag$ &	1.48	&	3.46	&	2.10E-12	&	-1.10	&	2.51	&	3.604	&	3.606	&	4.64	&	4.63	\\
\hline
\end{tabular}
\end{footnotesize}
\end{table*}


\begin{table*}
\begin{footnotesize}
\begin{tabular}{ l l l p{1.7cm} l l p{1.7cm} p{1.7cm} p{1.7cm} p{1.7cm}}\hline
HD	                       &	$B-V$	&	$V-K$	&	$\FMG$ 	&  $BC_{\rm V}$	&  $BC_{\rm K}$	& $\log{T_{\rm eff}}$ &	$\log{T_{\rm eff}}$ & $\FMGL$    & $\FMGL$	\\
...	                       &	...	&	...	&	...    	&  ...   	&  ...   	& from $B-V$          & from $V-K$          & from $B-V$ & from $V-K$	\\ \hline \hline
{\1p}61935	                 &	1.01	&	2.26	&	5.95E-12	&	-0.39	&	1.97	&	3.684	&	3.686	&	5.28	&	5.31	\\
{\1p}62044 *	           &	1.11	&	2.46	&	1.29E-10	&	-0.51	&	2.10	&	3.667	&	3.667	&	6.62	&	6.66	\\
{\1p}62345	                 &	0.92	&	2.00	&	9.18E-12	&	-0.30	&	1.77	&	3.699	&	3.716	&	5.42	&	5.49	\\
{\1p}62509 *	           &	0.99	&	2.09	&	9.94E-11	&	-0.37	&	1.84	&	3.687	&	3.705	&	5.41	&	5.53	\\
{\1p}63032	                 &	1.57	&	3.59	&	2.24E-11	&	-1.37	&	2.55	&	3.586	&	3.600	&	4.94	&	4.98	\\
{\1p}63700 *	           &	1.08	&	2.04	&	3.32E-11	&	-0.42	&	1.80	&	3.680	&	3.711	&	5.76	&	5.81	\\
{\1p}69267	                 &	1.45	&	3.26	&	1.35E-11	&	-1.05	&	2.45	&	3.608	&	3.615	&	4.90	&	5.00	\\
{\1p}71369 $\dagger$ $\ddag$ &	0.84	&	1.96	&	1.04E-11	&	-0.23	&	1.73	&	3.715	&	3.721	&	5.47	&	5.48	\\
{\1p}73974	                 &	0.87	&	1.96	&	1.44E-12	&	-0.26	&	1.73	&	3.709	&	3.721	&	6.00	&	5.96	\\
{\1p}76294	                 &	0.96	&	2.37	&	1.31E-11	&	-0.35	&	2.04	&	3.690	&	3.675	&	5.33	&	5.26	\\
{\1p}77912	                 &	0.97	&	1.97	&	9.19E-12	&	-0.30	&	1.74	&	3.700	&	3.720	&	5.82	&	5.85	\\
{\1p}78647	                 &	1.61	&	3.78	&	1.01E-10	&	-1.60	&	2.60	&	3.574	&	3.593	&	4.90	&	5.08	\\
{\1p}80493	                 &	1.53	&	3.73	&	2.25E-11	&	-1.24	&	2.59	&	3.594	&	3.594	&	4.84	&	4.85	\\
{\1p}81797 $\dagger$	     &	1.42	&	3.07	&	3.93E-11	&	-0.98	&	2.38	&	3.613	&	3.625	&	4.80	&	4.95	\\
{\1p}81817	                 &	1.38	&	3.20	&	6.33E-12	&	-0.84	&	2.43	&	3.626	&	3.618	&	5.03	&	4.91	\\
{\1p}82210	                 &	0.77	&	2.00	&	3.05E-11	&	-0.18	&	1.77	&	3.727	&	3.716	&	6.48	&	6.41	\\
{\1p}82668 $\dagger$	     &	1.51	&	3.56	&	1.48E-11	&	-1.21	&	2.54	&	3.596	&	3.601	&	4.68	&	4.75	\\
{\1p}84441	                 &	0.78	&	1.68	&	3.73E-11	&	-0.19	&	1.43	&	3.725	&	3.766	&	5.93	&	6.05	\\
{\1p}85503	                 &	1.21	&	2.48	&	8.19E-12	&	-0.63	&	2.11	&	3.650	&	3.665	&	5.16	&	5.31	\\
{\1p}88284 $\ddag$	     &	1.00	&	2.05	&	6.14E-12	&	-0.38	&	1.81	&	3.685	&	3.709	&	5.17	&	5.31	\\
{\1p}89388	                 &	1.47	&	3.13	&	2.41E-11	&	-1.02	&	2.40	&	3.610	&	3.621	&	5.12	&	5.20	\\
{\1p}89484	                 &	1.12	&	2.78	&	4.07E-11	&	-0.52	&	2.26	&	3.665	&	3.642	&	5.22	&	5.11	\\
{\1p}89758	                 &	1.58	&	4.00	&	3.42E-11	&	-1.44	&	2.65	&	3.582	&	3.585	&	4.86	&	4.88	\\
{\1p}90610	                 &	1.39	&	3.19	&	6.80E-12	&	-0.92	&	2.43	&	3.618	&	3.618	&	5.00	&	5.02	\\
{\1p}93497	                 &	0.89	&	2.04	&	8.41E-11	&	-0.28	&	1.80	&	3.705	&	3.711	&	6.06	&	6.08	\\
{\1p}93813	                 &	1.22	&	2.82	&	1.56E-11	&	-0.64	&	2.28	&	3.648	&	3.640	&	5.12	&	5.11	\\
{\1p}94264	                 &	1.03	&	2.23	&	3.22E-12	&	-0.43	&	1.95	&	3.678	&	3.689	&	4.92	&	5.01	\\
{\1p}95272                   &	1.06	&	2.30	&	5.79E-12	&	-0.46	&	2.00	&	3.673	&	3.682	&	5.25	&	5.33	\\
{\1p}95689 $\ddag$	     &	1.05	&	2.60	&	3.64E-11	&	-0.44	&	2.18	&	3.677	&	3.655	&	5.17	&	5.07	\\
{\1p}96833 $\dagger$ $\ddag$ &	1.13	&	2.53	&	1.11E-11	&	-0.53	&	2.14	&	3.663	&	3.661	&	5.04	&	5.07	\\
{\1p}98262                   &	1.36	&	3.09	&	1.59E-11	&	-0.79	&	2.39	&	3.631	&	3.624	&	5.16	&	5.11	\\
{\1p}98430	                 &	1.09	&	2.64	&	1.17E-11	&	-0.49	&	2.20	&	3.668	&	3.652	&	5.32	&	5.25	\\
102350	                 &	0.85	&	1.74	&	1.02E-11	&	-0.25	&	1.50	&	3.711	&	3.756	&	5.74	&	5.88	\\
104979 $\dagger$	           &	0.95	&	2.06	&	5.11E-12	&	-0.34	&	1.82	&	3.692	&	3.709	&	5.34	&	5.43	\\
106677	                 &	1.10	&	2.56	&	1.59E-11	&	-0.51	&	2.15	&	3.667	&	3.659	&	6.53	&	6.49	\\
107328	                 &	1.14	&	2.83	&	4.60E-12	&	-0.55	&	2.28	&	3.660	&	3.640	&	5.42	&	5.31	\\
107446	                 &	1.37	&	3.22	&	1.11E-11	&	-0.87	&	2.44	&	3.623	&	3.617	&	4.98	&	4.96	\\
108903 *	                 &	1.59	&	4.82	&	1.33E-10	&	-1.54	&	2.80	&	3.577	&	3.563	&	4.80	&	4.54	\\
108907 *	                 &	1.56	&	4.40	&	6.26E-12	&	-1.39	&	2.73	&	3.585	&	3.573	&	4.93	&	4.71	\\
109379 $\dagger$ $\ddag$     &	0.88	&	1.89	&	2.36E-11	&	-0.27	&	1.67	&	3.707	&	3.731	&	5.50	&	5.60	\\
111812	                 &	0.65	&	1.58	&	1.39E-11	&	-0.10	&	1.30	&	3.755	&	3.785	&	6.45	&	6.46	\\
112300	                 &	1.55	&	4.52	&	2.45E-11	&	-1.35	&	2.75	&	3.587	&	3.570	&	4.90	&	4.65	\\
113226	                 &	0.92	&	2.16	&	2.17E-11	&	-0.31	&	1.90	&	3.698	&	3.697	&	5.49	&	5.50	\\
113996	                 &	1.45	&	3.22	&	3.36E-12	&	-1.05	&	2.44	&	3.608	&	3.617	&	4.81	&	4.92	\\
115659 $\dagger$ $\ddag$     &	0.91	&	1.93	&	1.62E-11	&	-0.30	&	1.70	&	3.701	&	3.726	&	5.44	&	5.55	\\
116204	                 &	1.12	&	2.83	&	8.19E-12	&	-0.53	&	2.28	&	3.663	&	3.639	&	6.63	&	6.49	\\
123139	                 &	1.00	&	2.32	&	3.48E-11	&	-0.39	&	2.01	&	3.684	&	3.680	&	5.29	&	5.31	\\
124897 *	                 &	1.24	&	2.85	&	5.06E-10	&	-0.67	&	2.29	&	3.645	&	3.638	&	5.35	&	5.36	\\
127665	                 &	1.28	&	2.77	&	1.03E-11	&	-0.74	&	2.26	&	3.637	&	3.643	&	5.04	&	5.14	\\
127700	                 &	1.40	&	3.34	&	7.37E-12	&	-0.92	&	2.48	&	3.618	&	3.611	&	5.02	&	4.98	\\
129078	                 &	1.39	&	3.06	&	3.96E-12	&	-0.92	&	2.38	&	3.618	&	3.625	&	4.59	&	4.66	\\
129456 $\dagger$ $\ddag$     &	1.34	&	3.22	&	5.00E-12	&	-0.81	&	2.44	&	3.628	&	3.617	&	4.86	&	4.80	\\
131873	                 &	1.45	&	3.32	&	5.10E-11	&	-1.05	&	2.47	&	3.608	&	3.612	&	4.90	&	4.98	\\
133208 $\dagger$ $\ddag$     &	0.93	&	2.21	&	8.73E-12	&	-0.32	&	1.93	&	3.696	&	3.692	&	5.34	&	5.32	\\
133216	                 &	1.65	&	4.57	&	3.49E-11	&	-1.89	&	2.76	&	3.560	&	3.569	&	4.68	&	4.71	\\
134505 $\dagger$	           &	0.91	&	2.12	&	1.03E-11	&	-0.30	&	1.86	&	3.701	&	3.702	&	5.41	&	5.42	\\
135722	                 &	0.95	&	2.20	&	1.23E-11	&	-0.33	&	1.93	&	3.694	&	3.692	&	5.47	&	5.47	\\
137759	                 &	1.16	&	2.59	&	1.28E-11	&	-0.56	&	2.17	&	3.658	&	3.656	&	5.18	&	5.22	\\
140573	                 &	1.16	&	2.46	&	1.94E-11	&	-0.56	&	2.10	&	3.658	&	3.667	&	5.10	&	5.20	\\
141714	                 &	0.78	&	1.88	&	2.03E-11	&	-0.18	&	1.65	&	3.727	&	3.733	&	6.33	&	6.32	\\
146051	                 &	1.57	&	3.86	&	3.71E-11	&	-1.39	&	2.62	&	3.585	&	3.590	&	4.79	&	4.86	\\
146791	                 &	0.96	&	2.05	&	1.27E-11	&	-0.34	&	1.81	&	3.692	&	3.709	&	5.38	&	5.47	\\
147675	                 &	0.91	&	2.09	&	2.18E-11	&	-0.30	&	1.85	&	3.701	&	3.704	&	5.92	&	5.93	\\
148387 $\dagger$	           &	0.90	&	2.13	&	1.99E-11	&	-0.30	&	1.87	&	3.701	&	3.700	&	5.43	&	5.43	\\
\hline
\end{tabular}
\end{footnotesize}
\end{table*}


\begin{table*}
\begin{footnotesize}
\begin{tabular}{ l l l p{1.7cm} l l p{1.7cm} p{1.7cm} p{1.7cm} p{1.7cm}}\hline
HD	                   &	$B-V$	&	$V-K$	&	$\FMG$ 	&  $BC_{\rm V}$	&  $BC_{\rm K}$	& $\log{T_{\rm eff}}$ &	$\log{T_{\rm eff}}$ & $\FMGL$    & $\FMGL$	\\
...	                   &	...	&	...	&	...    	&  ...   	&  ...   	& from $B-V$          & from $V-K$          & from $B-V$ & from $V-K$	\\ \hline \hline
148856	             &	0.93	&	2.01	&	2.19E-11	&	-0.32	&	1.78	&	3.696	&	3.714	&	5.46	&	5.55	\\
150798	             &	1.41	&	2.97	&	9.08E-11	&	-0.88	&	2.34	&	3.622	&	3.630	&	5.21	&	5.30	\\
150997 *	             &	0.90	&	2.12	&	2.43E-11	&	-0.30	&	1.86	&	3.701	&	3.702	&	5.81	&	5.82	\\
153210	             &	1.15	&	2.44	&	1.21E-11	&	-0.56	&	2.08	&	3.658	&	3.669	&	5.11	&	5.23	\\
153751	             &	0.86	&	1.92	&	2.70E-11	&	-0.26	&	1.69	&	3.709	&	3.727	&	6.20	&	6.24	\\
156283 $\ddag$	       &	1.40	&	3.08	&	1.23E-11	&	-0.88	&	2.39	&	3.622	&	3.624	&	4.84	&	4.88	\\
157244	             &	1.42	&	3.08	&	5.43E-11	&	-0.90	&	2.38	&	3.620	&	3.624	&	5.34	&	5.37	\\
157999	             &	1.36	&	2.92	&	7.51E-12	&	-0.80	&	2.32	&	3.630	&	3.634	&	5.16	&	5.13	\\
159181 *	             &	0.92	&	1.93	&	1.04E-10	&	-0.25	&	1.71	&	3.712	&	3.725	&	6.23	&	6.25	\\
161096	             &	1.16	&	2.30	&	1.76E-11	&	-0.56	&	2.00	&	3.658	&	3.682	&	5.11	&	5.30	\\
161892	             &	1.18	&	2.53	&	1.37E-11	&	-0.59	&	2.14	&	3.655	&	3.661	&	5.15	&	5.23	\\
163588	             &	1.17	&	2.65	&	9.24E-12	&	-0.58	&	2.20	&	3.657	&	3.651	&	5.20	&	5.22	\\
163770 *	             &	1.28	&	2.67	&	1.42E-11	&	-0.68	&	2.21	&	3.644	&	3.650	&	5.35	&	5.39	\\
163993	             &	0.92	&	2.17	&	2.20E-11	&	-0.31	&	1.91	&	3.698	&	3.695	&	5.84	&	5.83	\\
167618	             &	1.57	&	4.69	&	3.41E-11	&	-1.39	&	2.78	&	3.585	&	3.566	&	4.90	&	4.60	\\
168723	             &	0.93	&	2.16	&	1.22E-11	&	-0.32	&	1.90	&	3.696	&	3.696	&	5.38	&	5.40	\\
169414	             &	1.16	&	2.50	&	7.95E-12	&	-0.56	&	2.12	&	3.658	&	3.663	&	5.20	&	5.28	\\
169916	             &	1.02	&	2.47	&	2.01E-11	&	-0.40	&	2.10	&	3.682	&	3.666	&	5.35	&	5.29	\\
171443 $\dagger$	       &	1.30	&	2.95	&	6.71E-12	&	-0.75	&	2.34	&	3.635	&	3.631	&	4.96	&	4.98	\\
174974	             &	1.23	&	2.42	&	9.19E-12	&	-0.59	&	2.08	&	3.654	&	3.670	&	5.64	&	5.59	\\
177716	             &	1.16	&	2.83	&	1.44E-11	&	-0.56	&	2.28	&	3.658	&	3.639	&	5.24	&	5.16	\\
183492 $\dagger$ $\ddag$ &	1.02	&	2.26	&	1.24E-12	&	-0.41	&	1.97	&	3.680	&	3.686	&	5.22	&	5.26	\\
186791 *	             &	1.46	&	3.31	&	3.90E-11	&	-1.02	&	2.47	&	3.610	&	3.612	&	5.06	&	5.09	\\
187076	             &	1.27	&	4.12	&	7.40E-11	&	-0.65	&	2.68	&	3.647	&	3.581	&	6.02	&	5.39	\\
188650	             &	0.62	&	1.25	&	3.20E-12	&	-0.08	&	0.70	&	3.764	&	3.874	&	6.20	&	6.28	\\
188947	             &	1.01	&	2.48	&	6.51E-12	&	-0.39	&	2.11	&	3.684	&	3.665	&	5.30	&	5.22	\\
192876	             &	0.86	&	2.18	&	1.48E-11	&	-0.21	&	1.91	&	3.721	&	3.694	&	6.04	&	5.84	\\
196171	             &	0.99	&	2.38	&	1.36E-11	&	-0.37	&	2.05	&	3.687	&	3.674	&	5.33	&	5.28	\\
197989	             &	1.01	&	2.47	&	2.68E-11	&	-0.40	&	2.10	&	3.682	&	3.666	&	5.34	&	5.28	\\
198700	             &	1.19	&	2.40	&	1.87E-11	&	-0.55	&	2.06	&	3.660	&	3.672	&	5.51	&	5.58	\\
200905	             &	1.49	&	3.43	&	2.09E-11	&	-1.10	&	2.50	&	3.604	&	3.607	&	5.13	&	5.08	\\
202109 $\dagger$	       &	0.98	&	2.01	&	8.72E-12	&	-0.36	&	1.77	&	3.689	&	3.715	&	5.18	&	5.32	\\
203387	             &	0.87	&	2.01	&	1.98E-11	&	-0.26	&	1.78	&	3.709	&	3.714	&	6.09	&	6.10	\\
204075	             &	0.96	&	1.98	&	2.65E-11	&	-0.35	&	1.74	&	3.690	&	3.719	&	5.90	&	6.02	\\
205435	             &	0.87	&	2.04	&	2.58E-11	&	-0.27	&	1.81	&	3.707	&	3.710	&	6.07	&	6.09	\\
205478	             &	1.00	&	2.25	&	8.31E-12	&	-0.39	&	1.97	&	3.684	&	3.687	&	5.34	&	5.39	\\
206778	             &	1.45	&	3.05	&	9.58E-11	&	-0.99	&	2.38	&	3.612	&	3.626	&	5.34	&	5.44	\\
206859	             &	1.07	&	2.42	&	1.50E-11	&	-0.40	&	2.07	&	3.682	&	3.671	&	5.83	&	5.71	\\
206952	             &	1.09	&	2.78	&	2.10E-11	&	-0.49	&	2.26	&	3.668	&	3.643	&	5.97	&	5.84	\\
207089	             &	1.27	&	2.58	&	1.09E-11	&	-0.65	&	2.16	&	3.647	&	3.657	&	5.83	&	5.84	\\
208816	             &	0.72	&	2.78	&	3.32E-11	&	-0.12	&	2.26	&	3.747	&	3.643	&	6.85	&	5.35	\\
209750 *	             &	0.89	&	2.15	&	5.00E-11	&	-0.23	&	1.89	&	3.716	&	3.698	&	6.00	&	5.83	\\
210745	             &	1.49	&	2.84	&	3.07E-11	&	-1.07	&	2.29	&	3.606	&	3.638	&	5.19	&	5.44	\\
211388	             &	1.39	&	2.95	&	4.96E-12	&	-0.90	&	2.34	&	3.620	&	3.631	&	4.84	&	4.91	\\
211416	             &	1.37	&	3.19	&	1.76E-11	&	-0.88	&	2.43	&	3.622	&	3.618	&	6.28	&	4.89	\\
213080	             &	1.54	&	4.94	&	1.43E-11	&	-1.28	&	2.82	&	3.592	&	3.560	&	6.53	&	4.51	\\
216228	             &	1.04	&	2.19	&	1.02E-11	&	-0.44	&	1.92	&	3.677	&	3.693	&	6.47	&	5.41	\\
216386	             &	1.59	&	4.29	&	1.58E-11	&	-1.49	&	2.71	&	3.579	&	3.576	&	6.34	&	4.67	\\
216489	             &	1.10	&	2.52	&	4.17E-11	&	-0.51	&	2.13	&	3.667	&	3.662	&	8.00	&	6.77	\\
217906	             &	1.64	&	4.76	&	5.30E-11	&	-1.80	&	2.79	&	3.564	&	3.564	&	6.22	&	4.50	\\
218356 *	             &	1.23	&	2.85	&	4.86E-11	&	-0.67	&	2.29	&	3.645	&	3.638	&	7.57	&	6.21	\\
219615	             &	0.90	&	2.27	&	7.61E-12	&	-0.30	&	1.98	&	3.701	&	3.685	&	6.48	&	5.32	\\
224427	             &	1.54	&	4.57	&	7.40E-12	&	-1.31	&	2.76	&	3.589	&	3.569	&	6.44	&	4.57	\\
\hline
\end{tabular}
\end{footnotesize}
\end{table*}


\section{Deriving the Mg~II {\ithk} chromospheric emission line
         surface flux}

\subsection{Measurements of the Mg~II {\ithk} chromospheric 
emission using the IUE final data archive}

The Mg~II lines, representing a transition to the Mg~II ground level,
form a strong doublet, denoted as Mg~II~{\it h} (2803 {\AA}) and
Mg~II~{\it k} (2796 {\AA}), respectively.  For our measurements of the
chromospheric Mg II emission line fluxes, only well-exposed spectra taken
in the high resolution mode are suitable.  There is a substantial body 
of previous work on the Mg~II emission in late-type giants and supergiants.
Examples include papers by Stencel et al. (1980), Simon \& Drake (1989),
Dupree, Hartmann \& Smith (1990) and Peterson \& Schrijver (1997).
These authors obtained and analyzed IUE spectra for various samples
of evolved stars, including metal-deficient stars.  For many of
these stars, they were able to identify asymmetric Mg~II profiles,
which were considered a unique signature of differentially expanding
atmospheres due to the presence of outward mass motions.
In the view of the evolutionary status of these stars, the authors
often attribute this finding to the action of hydrodynamic processes,
i.e., acoustic or pulsational waves; see, e.g., discussion by
Dupree et al. (1990).

The final IUE data archive (Nichols \& Linsky 1996; Nichols 1998) is,
thanks to the 19 years of record-lifetime of the venerable IUE satellite 
(i.e., IUE accomplished over 100,000 observations between 1978 to 1996),
even today a highly valuable data base, especially for stellar research.
IUE spectra have, in their high-resolution mode, a spectral resolution 
of 20,000 and cover the wavelength range of 3400 to 1150~{\AA}.  IUE had
two individual echelle spectrographs equipped with two cameras each, the
primary (P) and redundant (R) cameras. Spectra of the longer wavelength 
range (LW), which included the chromospheric Mg~II doublet, were taken 
by the LWR or the LWP camera.

For our study, we consider Mg~II {\ithk} IUE observations for a
total of 177 cool giants and supergiants of spectral types G, K, and M,
encompassing the luminosity classes III to I, of which surface fluxes can be 
derived by the means described in Sect. 2.2. This heterogeneous 
stellar sample covers a conveniently wide, two-dimensional array of 
physical parameters (i.e., $T_{\rm eff}$ and $g$), as well as different 
evolutionary stages (see Sect. 4). A total of 16 objects have been observed
several times and in many cases showed real variability.  Mostly, the variability
did not change the line profiles.  In cases of multiple observations, we focused
on the lowest flux measurement for each star since it is expected
to be closest to the basal flux, see Sect. 3.

The IUE database is available online 
from \verb|http://archive.stsci.edu/iue/search.php|.
Note that the photospheric UV flux in such cool stars is very low or
virtually absent compared to the Mg~II {\ithk} emission line peaks.
This makes the magnesium doublet highly favourable 
for chromospheric observational research.  
As mentioned above, another big advantage of the Mg~II emission 
lines for chromospheric studies is that their peak emissivity, in essence, 
covers a wide range of the chromospheric temperature distribution.  This not 
only invokes a very strong chromospheric emission line output, second only 
to the hydrogen lines, but even more importantly, the Mg~II emission flux 
is able to serve as a universal probe of the chromospheric energy dissipation 
and radiative losses, 
as pointed out a long time ago (e.g., Linsky \& Ayres 1978, see inset of
their Fig.~1). Since the radiative cooling is still dominated by hydrogen,
there is little, if any dependence of the Mg~II line flux on metallicity 
(Cardini 2005).

Like most other chromospheric studies, our measurements were made 
for both Mg II {\ithk} lines. A simple integration over the line profile
was made after subtracting the photospheric flux; see Fig. 1 for an example
of Mg~II {\it k} with the shaded area corresponding to the line integration.
We also determined the photospheric flux at 2795.5 {\AA}.  This simple line
integration procedure yields the physical emission line fluxes at Earth,
$F_{\rm Mg II-IUE}$, for each of our sample stars (see Table 1).
Stars, for which the lowest measurement was used from multiple observations,
are marked with an asterisk ($\ast$).  Stars used to determine the basal flux
line through a statistical fitting procedure (see Sect. 3.1) are marked with
a dagger ($\dagger$) or double dagger ($\ddag$), respectively,
depending on whether the $B-V$ or $V-K$ relation was employed for
the color transformation (see Sect. 2.2).


\subsection{Deriving Mg II emission line surface fluxes}

A more difficult step consists in deriving the stellar chromospheric surface fluxes 
$\FMG$. This could be done by directly using the distances and radii of the
sample stars. In fact, direct parallax measurements are mostly quite accurate, 
i.e., almost 80\% of the giants in our sample have a measurement error of only 
10\% or less of its parallax. However, the derivation of the radii still require the 
bolometric correction $BC$ of each star. In this respect, direct determinations 
of the surface fluxes do not suffer any less from systematic errors than
the various parallax-independent approaches, beginning with the Barnes-Evans 
relation for the visual surface brightness (Barnes, Evans \& Moffett 1978). 
Deriving the angular diameters and, hence, the observed fluxes over surface flux 
ratios this way will depend on a color relation. 

Here we use the relation by Oranje, Zwaan \& Middelkoop (1982) that uses $BC$
\begin{equation}
\log{\frac{F}{f}} = 0.35 + 4\log{T_{\rm eff}} + 0.4 (V+BC) \ .
\label{eq:SSF}
\end{equation}
The constant 0.35 has been adjusted to be consistent with the 
revised solar quantities by Cox (2001).

The required values for $T_{\rm eff}$, $BC_{\rm V}$ and $BC_{\rm K}$ were 
then derived from the color indices of each giant by the $B-V$ and $V-K$ color, 
using the color relations from Buzzoni et al. (2010):
\begin{equation}
 B-V=1.906\left[ BC^{2}_{\rm V} \exp(BC_{\rm V})\right]^{0.3}
\end{equation}
\begin{equation}
 BC_{\rm V}=-{\rm exp}(27500~{\rm K}/T_{{\rm eff}})/1000
\end{equation}

\begin{equation}
 V-K = 1 / (1 - 0.283~BC_{\rm K})
\end{equation}
\begin{equation}
 BC_{\rm K} = -6.75 \log{T_{\rm eff}/9500~{\rm K}}
\end{equation}

We obtained the individual $V-K$ values from the Two Micron All Sky Survey 
(2MASS). The $V-K$ color index is considered to be much more temperature 
sensitive than the $B-V$ color, especially for evolved stars, 
but there is a trade-off:  Extinction in $V-K$ reaches 91\% of the visual 
interstellar absorption $A_v$ (see Whittet \& van Breda 1978), while 
$B-V$ is well known to amount only to 32\% of $A_v$. For this reason, and 
for an appropriate comparison with historic work on the basal flux line, which 
relied mostly on $B-V$ color relations, we use here the $V-K$ derived data-set
only for estimating the magnitude of the systematic errors (see below) inherent
in the above relations.

Noting that our sample consists of giants stars, which typically
are much more distant than main-sequence stars, interstellar absorption and 
extinction does become an issue. Hence, both color indices, $B-V$ and $V-K$, 
were individually corrected according to an average, distance-dependent 
interstellar absorption $A_v$ of 1~mag per kpc. The distances were 
obtained from the parallaxes given by the Hipparcos catalogue. Of course, 
this correction only avoids systematic errors for the sample; however, it
is not very accurate for the individual objects. 

Figure 2 shows the derived chromospheric Mg~II {\ithk} line 
emission surface fluxes of the 177 cool giants and supergiants 
obtained from the IUE archive, to which the above $B-V$ relation is applied,
as a function of the effective temperature (see Sect. 3 and Table 1).
Stars with multiple observations and clearly variable Mg~II fluxes have been 
shown with their minimum fluxes. Thanks to the richness of the sample, the 
distribution of chromopsheric fluxes shows, within some scatter, a lower 
limit with a simple dependence on $T_{\rm eff}$ (i.e., linear in double
logarithmic representation).


\section{The basal chromospheric flux as a function of $T_{\rm eff}$}

\subsection{Detailed statistical analysis}

In order to gauge the significance of the line associated with the
lower limits of the Mg~II chromospheric emission fluxes (see Fig. 2), 
we pursue a detailed statistical analysis of the data.
In principle, we distinguish between {\it systematic uncertainties}
associated with the uncertainties of the individual data
and {\it statistical uncertainties}, due to the uncertainty of the
basal flux line itself (i.e., slope and $y$-axis intercept), associated with
the spread of the observational data.

As data uncertainties, we take the estimated double amplitude of 
20\%, i.e., $\pm 0.04$ in $\FMGL$.  The value of 20\% is
essentially completely due to the $F$ and $f$ values (see Eq.~1),
i.e., by the absolute flux measurements for Mg~II {\ithk} and the
conversion to the stellar surface fluxes.  Physical quantities affecting
the results include the visual magnitude of the star $V$, the
bolometric correction $BC$ as well as the adopted stellar distances
and the stellar radii.  A further small, almost negligible
contribution to the systematic uncertainty arises from the subtraction
of the stellar continuum from the flux measurements (see Fig.~1).
This latter uncertainty is assumed to be
negligible owing to the diminutive contribution of the
the continuum to the measured Mg~II {\it h} and {\it k} lines.

As part of our study, we deduced two different lines for the
lower limits of the Mg~II fluxes in response to the usage of
either the $B-V$ or $V-K$ relation for the color transformation
(see Sect. 2.2).  In the first case, the basal flux line is determined
by using $N_* = 22$ stars, whereas in the second case $N_* = 15$ stars
were used.  The objects
are listed in Table 1, noting that the selected objects for
the two determinations are labelled by a dagger
($\dagger$) or double dagger ($\ddag$), respectively.

The lines given in the form
\begin{equation}
{\FMGL} = B \log{T_{\rm eff}} + A
\end{equation}
are calculated by invoking a least-square fit with consideration
of the uncertainty bar for its individual data point.  In doing so,
we used the statistical fitting procedure ({\tt SUBROUTINE FIT}) as
described by Press et al. (1986); see Table 2 for our results.
This procedure also gives information on the uncertainty (i.e.,
standard deviation) concerning the slope ($\sigma_{\rm B}$) and
the $y$-axis intercept ($\sigma_{\rm A}$).  

\begin{table}
\caption{Statistical parameters of the basal flux line. } 
\begin{small}
\begin{tabular}{l c c c c c c}
\hline 
Model	& $N_*$ & $A$   &  $\sigma_{\rm A}$	&   $B$ &  $\sigma_{\rm B}$ &  $R_{\rm Sp}$  \\
\hline\hline
 $B-V$       & 22    &  -21.75  &  1.72 &  7.33  &  0.47  &  0.99995 \\
 $V-K$       & 15    &  -19.74  &  1.40 &  6.78  &  0.38  &  0.99989 \\
\hline
\end{tabular}
\end{small}
\end{table}

For later depiction, we also calculated the interval estimates
associated with the position of each basal flux line given by the
sample (see greyish area); see, e.g., Montgomery, Peck \& Vining (2006)
for a brief description of the involved algebra.  We considered
both the systematic uncertainty of the data and the statistical
uncertainty with respect to the two previously determined basal flux
lines.  We assumed a confidence interval (CI) of 95\%, corresponding
to 2.101 standard deviations.  The systematic and the statistical
uncertainties have been combined in a linear manner.  The size of
the uncertainties obtained for both $A$ and $B$ (see Table 2)
are mostly due to the relative small number of objects considered for
our analysis.  Fortunately, however, they do not result in an overly
large interval estimate for the position of either basal flux line.

A further result obtained by utilizing the procedure by Press et al.
(1986) consists in the computation of the Spearman rank-order coefficient
$R_{\rm Sp}$ (see Table 2) for the correlation between $\FMGL$ and
$\log{T_{\rm eff}}$.  An outcome of $-1$ would indicate a perfect
anti-correlation, whereas an outcome of $+1$ would indicate a perfect
(positive) correlation.  For the basal flux line obtained by employing
the $B-V$ method, we obtain a $R_{\rm Sp}$ value of 0.99995, whereas for
the basal flux line obtained by employing the $V-K$ method a $R_{\rm Sp}$
value of 0.99989 is found.  Thus, in conclusion, there is an almost perfect
(positive) correlation between $\FMGL$ and $\log{T_{\rm eff}}$ for both lines,
as expected.

Based on the above analysis, while noting that we prefer the $B-V$ based
relation (see Sect. 2.2), the empirical basal flux of the full sample can 
be represented by a straight line given as
\begin{equation}
\FMGL \ = \ 7.33 \log{T_{\rm eff}} - 21.75
\end{equation}
that is depicted in both Fig.~2 and Fig.~3.  It is also depicted when
comparing our results with previous empirical results as well as results
from detailed model computations (see Fig.~4).

Here, the statistical uncertainty of the empirical slope of the basal flux 
line in the $\log{T_{\rm eff}}$ --- $\log{F}$ diagram, given as 7.33, is 0.47.
The systematic errors are of similar order.  When using the $V-K$ relations
instead, the slope is 6.78, and if furthermore the Barnes-Evans relation 
is used for deriving the surface fluxes, with $BC$ and $T_{\rm eff}$ 
taken from the $B-V$ relations above, the slope is 7.62. Hence, these
alternative values fall on either side of the purely $B-V$ based result above.

\begin{figure}
\centering
\begin{tabular}{c}
\epsfig{file=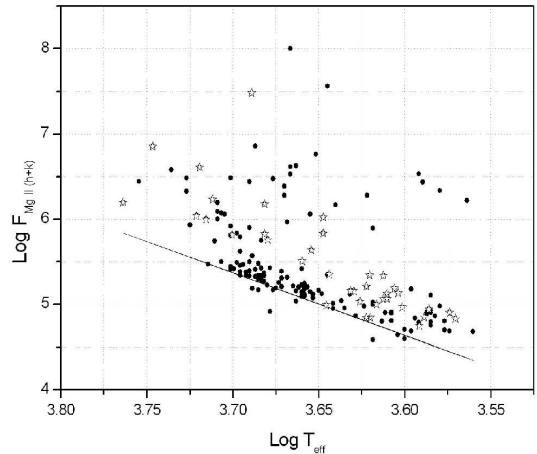,width=1.09\linewidth}
\end{tabular}
\caption{Measured Mg~II {\ithk} chromospheric emission line surface fluxes
         of 177 giants as a function of $T_{\rm eff}$ based on a $B-V$ relation
         (see text for details).
         Open star symbols represent giants of LC I, dark dots LCs II and III.
         The straight line represents the derived basal flux.  }
\label{}
\end{figure}

\begin{figure}
\centering
\begin{tabular}{c}
\epsfig{file=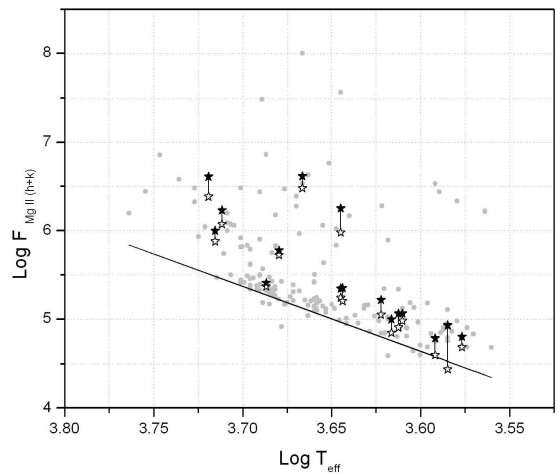,width=1.13\linewidth}
\end{tabular}
\caption{A plot of the 16 giants with multiple observations, 
         showing both their average and their minimum Mg~II {\ithk}
         fluxes, against the shaded backdrop of the total sample 
         as in Fig.~2.
}
\label{}
\end{figure}


\subsection{Is stellar variability consistent with the basal flux limit?}

Our sample consists of 16 giants and supergiants for which multiple 
observations exist (marked by an asterisk in Table 1) 
and to which the relations concerning the derivation of the Mg~II fluxes
(see Sect. 2.2) are applicable.  In most cases, these observations span over
at least several years, in some cases nearly the whole two decades
of the IUE life time, a timescale comparable to the expected duration of
long stellar activity cycles (e.g., Baliunas et al. 1995).

Hence, in view of all sporadic variations of stellar activity on timescales 
of months, as well as their systematic variations over the full activity cycle,
the range of this variability should clearly lie above the basal flux limit.
However, it should be noted that the large-scale of giant granular structure 
is also expected to yield a certain level of flux variation, expected to be
statistical in nature, even in the case of pure acoustic wave heating
(e.g., Judge \& Cuntz 1993).

Figure 3 indicates that this is indeed the case concerning the 16 stars.
It is noteworthy that, despite looking at
16 different objects over a relatively long period of time, {\it none} of
them showed a minimum Mg~II emission line flux dipping {\it below} the limit
given by the basal chromospheric flux.  Note that this assessment
is valid for both basal flux limits obtained by employing either the
$B-V$ or $V-K$ relationship; see Sect. 2.2.


\subsection{Comparison with previous results}

About the same temperature dependence as given above has been found in
earlier studies (Oranje \& Zwaan 1985; Schrijver 1987; Rutten et al. 1991),
which were however mostly aimed at MS stars.  For example, Rutten et al. (1991)
derived a dependence on the effective temperature akin to $T_{\rm eff}^8$. 
A paper focused on the basal flux in G and K-type giants has been
given by Strassmeier et al. (1994).  They measured the Ca~II line
emission using ground-based spectroscopy.  For the K-line they found a
basal flux represented by
\begin{equation}
\log{F_{\rm Ca~II}} = 8\log{T_{\rm eff}} - 24.8 \ ,
\end{equation}
which has a similar slope as the relationships obtained in our
present study, particularly the relation based on $B-V$ color 
transformation (see Eq.~7).

A more recent result for the emission flux in Ca~II was given by
Pasquini, de Medeiros \& Girardi (2000).  They obtained Ca~II H+K
high resolution observations for 60 evolved stars
of spectral type F to K located in five open clusters.  
Pasquini et al. (2000) found that
the Ca~II K fluxes scaled, in essence, linearly with the stellar rotational
velocity, whereas the basal flux line scaled with respect to the stellar
effective temperature according to 
$F_{\rm Ca II K} \propto T_{\rm eff}^{7.7}$. 

Hence, the slopes of the basal Mg~II flux emission (based on $B-V$ and
$V-K$ colors) as well as the slope of the previously derived basal Ca~II
flux emission fully agree with respect to each other within their
respective uncertainties, if available.   In absolute terms, the
Mg~II basal flux derived in the present study lies about 0.6 dex higher,
while the expected off-set should be only about 0.3 dex due to 
the intrinsically smaller emission measure of the Ca~II K lines compared to
the Mg~II lines, as well as other differences in the Ca~II and Mg~II line 
formation processes (for a detailed study see, e.g., Linsky \& Ayres 1978 and 
Ayres 1979). The remaining difference of a 0.3 dex higher basal flux
deduced in the present study is attributable to the fact that the earlier 
work did not fully consider underlying uncertainties.  The omission of
uncertainties readily results in an artificially broadened flux distribution
towards smaller values.  If we choose not to average over various borderline
cases (see Sect.~3.1) but simply set the basal flux line to consistently 
undercut {\it all} lowest flux values, our findings would essentially be in  
perfect agreement with the historical values.


\section{Stellar evolutionary aspects concerning the Mg~II emission}

Links to theoretical studies of angular momentum evolution have been
explored by Gray (1991), Schrijver (1993), Schrijver \& Pols (1993),
Charbonneau, Schrijver \& MacGregor (1997), and others.
These studies show that when
solar-type stars evolve away from the main-sequence, their rotation
slows down beyond what must be expected from the increase in the 
moment of inertia caused by changes in the internal mass distribution.
Hence, their angular momentum subsides, due to magnetic braking 
resulting from the onset of massive stellar winds. Thus, one would
rather expect little magnetic activity in evolved stars. But as
demonstrated by Fig.~2, this is mostly true only for M and late 
K giants (i.e., with $\log{T_{\rm eff}}<3.64$). By contrast, 
flux measurements for many G giants and G supergiants  
are spread over several orders of magnitude above their basal fluxes. 

A look at the evolutionary history of the respective giants shows that 
the latter group should indeed be expected to exhibit strong magnetic 
activity, whereas the former should not.  The same picture has been found 
empirically from the strength of coronal X-ray emission, and was explained
based on the same reasoning (Schr\"oder, H\"unsch \& Schmitt 1998):
G giants are usually first-time Hertzsprung-gap crossers of about 
1.5 to 2 $M_{\odot}$, whereas G supergiants are relatively massive (4 to
10 $M_{\odot}$) stars in their central He-burning phase.  All
these giants once were, as MS stars, too hot (i.e., spectral type earlier
than F0) to possess convective outer layers; those only
developed later. Hence, G giants and supergiants were unable to host a
solar-like magnetic dynamo until only very recently in their evolutionary
history, and their present magnetic activity has not suffered much 
from magnetic breaking. 

By contrast, K and M giants have mostly evolved from 
less massive MS stars, as K giant clump stars and as RGB/AGB stars.  Their 
progenitors have been active during their entire MS life times and suffered from 
magnetic breaking all along.  Hence, for these stars, like in regards to
their coronal X-ray emission, we find a reduced (but still not vanished) level
of magnetic activity.
It has therefore been argued (e.g., Dupree et al. 1990; Buchholz et al. 1998)
that the dominant chromospheric heating mechanism in single inactive
late-type (super-) giants might be essentially acoustic rather than
magnetic in nature.  Early results in support of this picture based
on the analysis of the evolution of rotation rates and chromospheric
activity of giants have been given by Rutten \& Pylyser (1988).

\begin{figure*}
\centering
\begin{tabular}{c}
\epsfig{file=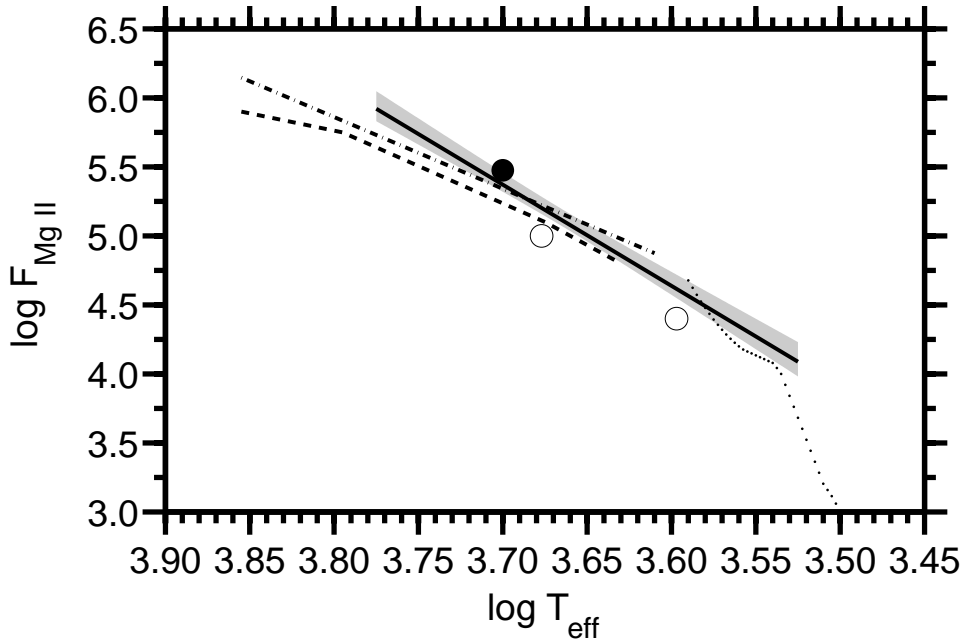,width=0.75\linewidth}
\end{tabular}
\caption{Diagram showing the basal flux limit for Mg~II {\ithk} as a
function of the stellar effective temperature obtained from $B-V$ 
colors (see Sect. 2) together with
results from previous observations and theoretical simulations.
The solid line shows our results for the object-selected basal flux line.
The uncertainties due to statistical and systematic observational effects
(see Sect. 3.4) are indicated by the greyish area.  The dashed line shows
the results from Rutten et al. (1991), whereas the dot-dashed line indicates
the earlier results from Schrijver (1987).  The dotted line shows the
Mg~II basal flux limit obtained through the sample by Judge \& Stencel (1991).
The circles represent results from previous theoretical simulations based
on acoustic energy dissipation given by Cuntz et al. (1994) (closed circle)
and Buchholz et al. (1998) (open circles).
}
\label{}
\end{figure*}


\section{Comparison with theoretical studies;
possible effects of the stellar surface gravity}

Figure 4 offers a detailed comparison of our results for the chromospheric
basal flux line with results from previous theoretical and empirical studies
as given by Schrijver (1987), Rutten et al. (1991),
Judge \& Stencel (1991), Cuntz et al. (1994), and Buchholz et al. (1998).
Even though the sample of stars presented here is much larger and has a different 
composition compared to some of the earlier studies, we find that our results 
are in good agreement, especially with the theoretical results obtained by
Buchholz et al. (1998).
With respect to the different approaches in the derivation of the 
chromospheric basal flux line and the associated systematic errors 
(see Sect. 3.1), no discernible differences were found.  The systematic
errors are commensurate to the uncertainties of the individual data.

As shown in Fig.~4, our observational results, i.e., the empirical basal flux
limits for the adopted sample including the biparabolic
envelopes for the associated statistical uncertainties, show considerable
agreement with the previous results by Schrijver (1987) and Rutten et al.
(1991).  Agreement is found for the relatively hot segment of the sample
($T_{\rm} \gta 4050$~K) with the previous results by Schrijver (1987)
and Rutten et al. (1991) as well as for the relatively cool segment of the
sample ($T_{\rm}\lta 3950$~K) with the result by Judge \& Stencel (1991);
note that none of these authors communicate any uncertainty bars
for their results.  Judge \& Stencel provided an empirical analysis
for the global thermodynamic properties of the outer atmospheres of giant
stars and a small number of supergiants by addressing both the mass-loss
energetics and the outer atmospheric heating in the framework of previously
proposed concepts.

Previous work by Linsky \& Ayres (1978), Schrijver (1987) and Rutten et al. 
(1991) revealed the surprising property that the Ca~II and Mg~II basal flux 
lines for MS stars and giants essentially coincide. This result is, in 
principle, also consistent with the findings from the more recent study by
Cardini (2005) based on 225 stars of luminosity class I to V, even though
it is providing some indications that the Mg~II {\it k} flux increases
slowly (i.e., by a factor of 2) with decreasing stellar gravity, especially
if supergiants are included.  On the other hand, it is highly noteworthy
that the targeted gravity range encompasses more than four orders of
magnitude and, furthermore, a factor of 2 is well within one standard
deviation of the position of the Mg~II basal flux line obtained by
the present study (see Fig.~4).

It has been suggested that the lack of a pronounced 
difference in the Mg~II (and Ca~II) basal flux limits is mainly caused by the
enormous differences in the geometrical extent of the stellar radiative
zones for the different types of stars, especially with respect to the
H$^-$ continuum (Ulmschneider 1988, 1989; Cuntz et al. 1994;
Buchholz et al. 1998).   In the framework of acoustic models, the
wave energy flux at photospheric levels is significantly increased
in giant stars compared to MS stars, but so is the extent and
efficiency of the H$^-$ radiative damping zones (Ulmschneider 1988, 1989).
However, both features largely off-set each other, leading to 
a very similar amount of available wave energy in the
Ca~II and Mg~II line formation region in MS and giant stars.
Consequently, for the special case of magnetically inactive stars,
this behaviour can readily explain a relatively similar basal flux emission
for both types of objects, consistent with observations.  Cuntz et al. (1994)
expanded this type of analysis to evolved stars of different metallicities
which led to the same type of outcome, thus providing consistency with
previous observational results by Dupree et al. (1990).


\section{Discussion and conclusions}

With an unprecedented sample size and time coverage, our study confirms
the concept of a basal chromospheric emission flux from a purely empirical point 
of view.  We confirm a strong dependence of the flux on $T_{\rm eff}$, while a
dependence on gravity (shift of the basal flux regarding samples of stars
of different luminosity classes) must be smaller than the involved statistical
errors.  In addition, all well documented cases of time-variable chromospheric
flux emission (see Sect. 3.2, Fig. 3) remain well above the basal flux limit.
We view this as further empirical evidence for a {\it physically different}
origin of the basal flux, which is obviously unrelated to the highly
time-variable heating processes associated with stellar magnetic activity.

We acknowledge that there is still an ongoing debate about the physical
origin of the basal flux.  Previously, Judge \& Cuntz (1993) confronted
the predictions of 1-D acoustic heating models for $\alpha$ Tauri (K5~III)
with HST-GHRS observations.  They arrived at a list of problems associated
with the disagreements between the theoretical acoustic models and the
observation.  Judge \& Cuntz (1993) concluded that $\alpha$ Tauri's
chromosphere might either (partially) be magnetically heated or
$\alpha$ Tauri has acoustic properties substantially different from
those of traditional mixing-length models or its chromosphere is
determined by 3-D turbulence and/or horizontal flows.  However, so far,
no detailed 3-D acoustic chromospheric heating models have been given for
giant or supergiant stars, which would allow contesting the empirically
deduced basal flux limits, although significant progress on the calculation
of 3-D non-magnetic chromosphere models has meanwhile been made for the Sun
(e.g., Wedemeyer et al. 2004).

We therefore suggest, also considering our discussion of Sect. 5, that the
processes for the heating and emission of chromospheric basal flux stars
are largely attributable to the dissipation of mechanical energy.  This form
of energy is inherent in any star and shows little variability with time
(except on stellar evolutionary time-scales).  As its physical manifestation,
acoustic waves and turbulent motions have been suggested as a candidate
mechanism long time ago (e.g., Narain \& Ulmschneider 1990).  In combination
they might also explain or contribute (in an appropriate 3-D implementation)
to the spectroscopically derived near-sonic (or apparently even supersonic)
turbulent velocities revealed by chromospheric line profiles.

The dissipation of a form of non-magnetic, supposedly acoustic wave energy
deposited in the turbulent chromosphere, agrees with the refined empirical
relationship, which we obtained for the basal flux including its dependence
on the stellar effective temperature.  Different approaches to derive the
surface fluxes and the values for the effective temperatures yielded
no significant differences (i.e., no more that 1$\sigma$ of the involved
statistical parameters).

In that sense, we can also confirm that our results agree with 
previously attained hydrodynamic chromosphere models, also noting that the
mechanical energy deposited in the Mg~II {\ithk} line formation regions
vary very little with gravity.  
In fact, within the uncertainties, our relation renders the same basal flux 
even over four orders of magnitude as has been attained for MS stars.  
Still, a small contribution to the basal flux from heating processes 
related to some permanent, weak magnetic carpet
cannot entirely be ruled out since the empirical basal flux appears
to be slightly higher (by about its inherent uncertainty) than the
fluxes suggested by previous acoustic heating models.
In fact, it has been argued that magnetic fields might also contribute
to balancing the basal emission flux (Judge \& Carpenter 1998),
although more detailed studies are still needed to further investigate
this claim.

It is also noteworthy that the upward propagating mechanical flux
and energy required for driving a cool wind emerging from
the outer chromosphere is about an order of 
magnitude smaller than the total mechanical turn-over of the turbulent 
chromosphere in the above mentioned models.  Hence, it may be possible
for a turbulent giant chromosphere to divert such a small fraction of its 
mechanical flux to drive a cool star wind.  It is intriguing that both
the basal flux and a matching wind energy input show about the same
dependency on $T_{\rm eff}$ (by a power of 7.5,
as assumed by Schr\"oder \& Cuntz 2005), and neither seems to depend
considerably on gravity.  An extended giant chromosphere
definitely assists a cool star wind in another aspect:  Mass loss emerging
from the outer chromosphere requires a lot less energy input compared to
a wind originating directly from the stellar photosphere due to the
reduction in the required amount of potential energy.
These two arguments strongly support the idea that the mechanical energy 
reservoir of the chromosphere, as probed by the basal flux, also nurtures 
the wind energy input, possibly associated with weak magnetic fields.

Updated models based on structured red giant winds have meanwhile been
presented by Suzuki (2007).  He attributes the occurrence of mass loss to
the action of Alfv\'en waves, excited by surface convection, which travel
outward while dissipating via non-linear processes to accelerate and heat
the stellar winds; see, e.g., Hartmann \& MacGregor (1980), Hartmann \&
Avrett (1984) and Airapetian et al. (2000) for previous models of mass loss
concerning low-gravity stars invoking Alfv\'en waves.
In case of hybrid stars, Suzuki (2007) advocates the existence of
hot bubbles, which are the source of the observed UV/soft X-ray emission.
Clearly, in the framework of cool giant chromospheres and winds, it appears
that there might be an intricate interplay of different processes operating on
different scales which are responsible for producing the observed phenomena
and features.

\section*{Acknowledgments}
This work has been supported by the Mexican Science Foundation 
CONACyT under their postgrad-student grant program and by CONACyT grant
No. 80804 (CB-2007). Furthermore, we are very grateful for the on-line 
availability of the IUE final data archive at STScI, without which this
study would not have been possible.

{}

\end{document}